\documentclass[twocolumn,
showpacs,nofootinbib,superscriptaddress]{revtex4}
\usepackage{dcolumn}
\usepackage{amsmath,amssymb,setspace,graphicx,subfigure,epsfig}
\usepackage{amsfonts}
\usepackage{latexsym}
\usepackage{epsfig}
\usepackage{color}

\begin{document}

\title{Update of axion CDM energy density}

\author{Kyu Jung Bae}
\email{baekj81@phya.snu.ac.kr}
\author{Ji-Haeng Huh}
\email{jhhuh@phya.snu.ac.kr}
\author{Jihn E. Kim}
\email{jekim@ctp.snu.ac.kr}
\affiliation{ Department of Physics and Astronomy and Center
for Theoretical Physics, Seoul National University, Seoul 151-747, Korea}

\begin{abstract}
We improve the estimate of the axion CDM energy density by considering the new values
of current quark masses, the QCD phase transition effect and a possible anharmonic effect.
\end{abstract}

\pacs{14.80.Mz, 12.38.Aw, 95.35.+d}
\preprint{SNUTP 08-003}

\maketitle

\section{Introduction}

The standard model is the greatest triumph of the 20th century particle physics. However, it suffers several naturalness problems. The so-called strong CP problem is one of them. Up to now, the most promising strong CP solution is axion \cite{PQ77}.  Although its coupling is very small, it could have affected the evolution of the Universe.

On the other hand, the observations like the galaxy rotation curve and the cosmic microwave background radiation tell us that a significant portion of the energy density of the universe cannot be explained by the ordinary baryonic matter. The recent analysis of WMAP data\cite{Komatsu:2008hk} combining the observation of the large scale structure prefers the standard vacuum energy and cold dark matter ($\Lambda$CDM) model, with the cold dark matter (CDM) density $\Omega_{\rm DM} h^2 \simeq 0.1143 \pm 0.0034$.

During the inflation, a sufficiently light scalar field becomes spatially homogeneous and its value is selected stochastically\cite{Linde82}. After the inflation, the Hubble parameter decreases as the universe expands.  If the Hubble parameter becomes smaller than some scale, usually the scalar mass scale, the scalar field starts to roll down. In the hydrodynamic description, this coherent oscillation works as  cold matter in the evolution of the universe. If the interaction is small enough to keep coherence until the matter-dominant era, it can be a part of the present matter energy density of the universe. The axion is that kind of particle and can be coherent until even now.

Just after the invisible axion was introduced \cite{Kim79}, people recognized its cosmological implications that the classical axion field oscillation behaves like CDM \cite{Preskill83}. On the axion CDM energy density, Turner presented a more accurate number including anharmonic effect and numerical instanton calculation in late 1980s \cite{Turner86}. DeGrand {\it et. al.} concerned about the possible effects of the chiral phase transition \cite{DeGrand:1985uq}.

After Turner's pioneering work, many input parameters in axion cosmology, especially the QCD parameters, are refined. We present the improved number for the axion dark matter relic density with these new parameters.  We also discuss the possible effects of the chiral phase transition on the total entropy and the axion number density. This occurs only when supercooling arises.

\section{computation of axion relic density}

Axion is a pseudo Nambu-Goldstone boson, acquiring mass from the instanton effect. The temperature dependence of the axion mass is important in calculating axion energy density for which we will use the result of Ref. \cite{Yaffe81} obtained in the dilute gas approximation. For the purpose of sketching the computation, we just parameterize the axion potential as
\begin{equation}
V(\theta)=-C(T)\cos(\theta),
\end{equation}
where $\theta=a/F_a$ and
\begin{equation}
C(T)=\alpha_{\rm inst}{\rm GeV}^4 (T_{\rm GeV})^{-n}
\end{equation}
where $T_{\rm GeV}=T/{\rm GeV}$.

The inflation freezes all fields whose masses are smaller than the Hubble parameter at that time. It also flattens the fluctuation of the field. Consequently, just after the inflation, the sufficiently light fields remain homogeneous with a certain value that is determined stochastically. The axion and the moduli are typical representatives of them.

At the end of inflation, the universe is reheated by some processes. Since axion is very weakly interacting, the reheated radiation of the early universe cannot destroy its coherence. As the universe cools down, the Hubble friction can no longer be the dominant term and the axion field starts rolling down the potential hill and continues to oscillate. Since the Hubble parameter becomes smaller and smaller than the axion mass, or the oscillating period of the axion is smaller than the Hubble time, its energy and pressure can be described by the time averages during each oscillation. In the case of the harmonic potential, the effective pressure is zero and the axion field acts like cold matter in the evolving universe. Moreover, if the axion can keep the coherence until the matter domination epoch, it can contribute to the present CDM energy density.

Since the axion mass depends on the temperature or the time, the comoving energy density conservation does not work in estimating the axion relic density. However, the adiabatic theorem says that as long as the adiabatic conditions $H,\dot m_a /m_a \ll m_a$ hold, we can find the {\it adiabatic invariant} $I$. For the harmonic potential, the invariant is the comoving axion number density $I=R^3 \rho_a / m_a(T)$. Since the coherent state is not the number eigenstate, the interpretation as the number density only works for the harmonic potential, i.e. the free field case. We will discuss the appropriate invariant quantity in the case of the interacting axion field.

\section{Temperature dependence of axion mass}

The QCD vacuum has the quark-gluon phase and the chiral symmetry breaking hadronic phase, separated by the critical temperature $T_{\rm c}$ of the chiral symmetry breaking.  $T_{\rm c}$ has been calculated to be $148^{+32}_{-31}(172^{+40}_{-34})$ MeV for three (two) light quarks \cite{Braun07}. This region is in the boundary of weak and strong coupling regimes and it is difficult to estimate the axion mass very accurately in this region. Early attempts to estimate the temperature dependent axion mass has been given in \cite{Steinhardt83,Turner86}. In these earlier studies, this phase transition was not considered and here we point out that the consideration of the phase transition does not lead to a substantial change in the estimation of the axion energy density.
\begin{figure}[!h]
\resizebox{0.95\columnwidth}{!}
{\includegraphics{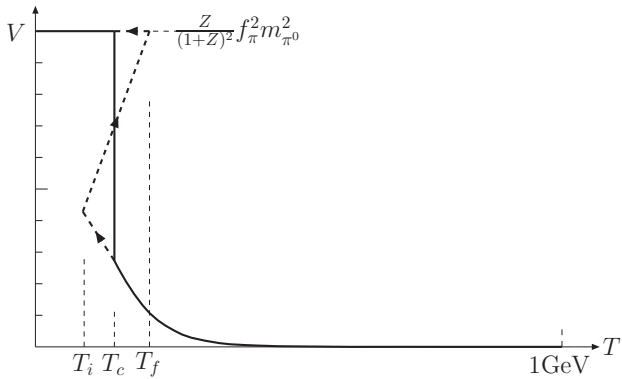}}
\caption{The phase transition near the critical temperature $T_{\rm c}\approx 150$ MeV. The solid line describes a smooth path first order transition and the dashed line describes a supercooling case.} \label{fig:axmass}
\end{figure}

\begin{figure}[!h]
\resizebox{0.95\columnwidth}{!}
{\includegraphics{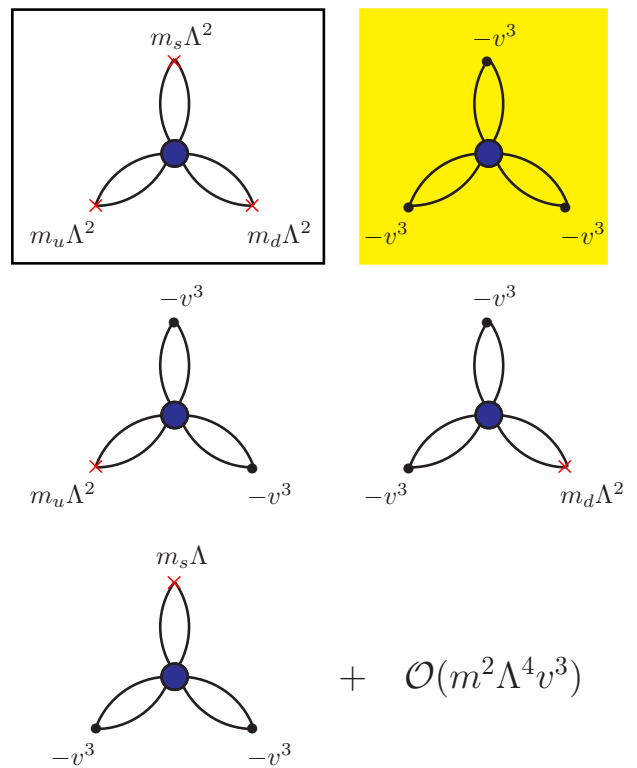}}
\caption{The determinental interaction of light quarks.}
\label{fig:tHooft}
\end{figure}

In the hadronic phase below the critical temperature, the correlation length of $q-\bar q$ pair is large and we assume a temperature independent height for the axion potential as schematically depicted in Fig. \ref{fig:axmass}
in the sudden phase change approximation. In this phase, the squared axion mass has the well-known one power of the light quark mass, and the coefficient of the axion potential is $f_\pi^2 m_\pi^2 Z/(1+Z)^2$ \cite{Baluni79} which is the second line in Fig. \ref{fig:tHooft} in the expansion of the determinental interaction \cite{tHooft76}.
On the other hand, in the quark-gluon phase mesons such as $\pi^0$ and $\eta'$ are not dynamical, and the first term of Fig. \ref{fig:tHooft} is the coefficient of the axion potential, which is $-K^{-5}\bar uu\bar dd \bar ss\cos\theta$ for three light quarks where $\theta$ is the QCD vacuum angle. Considering the $3\times 3$ mass matrix for $a,\pi^0,\eta'$, the insertions of $q\bar q$ VEVs (the lightly shaded diagram of Fig. \ref{fig:tHooft}) is the dominant contribution to the $\eta'$ mass \cite{tHooftU1} from which we estimate the strength $K^{-5}\approx m^2_{\eta'} f^2_{\eta'}/4v^9\simeq 10^{-13}{\rm MeV}^{-5}$ for $f_{\eta'}\simeq f_\pi$ in the hadronic phase and $v\approx 260$ MeV.  The same number $K^{-5}$ also appears in the quark-gluon phase. This value will be needed if the adiabatic approximation is violated, but we will show below that the adiabatic approximation is not violated in the bulk of the remaining parameter space and the number $K$ will not be used in the end.

Closing the quark lines to insert the current quark masses in the quark gluon phase, we obtain $-K^{-5}(m_u m_d m_s /\bar\rho^{6})\cos\theta$ where $\bar\rho$ is the effective instanton size in the instanton size integration at high temperature. Instead of this naive averaging, we use the original expression given in \cite{Yaffe81}
\begin{equation}
n(\rho,0)e^{\left\{-\frac13\lambda^2 (2N+N_f) -12 A(\lambda)\left[1+\frac16(N-N_f)\right]\right\} }
\label{InstantonCal}
\end{equation}
where $\lambda=\pi\rho T, A(\lambda)\simeq -\frac{1}{12}\ln (1+\lambda^2/3)+\alpha(1+\gamma\lambda^{-2/3})^{-8}$ with $\alpha=0.01289764$ and $\gamma=0.15858$. The zero temperature density is
\begin{eqnarray}
n(\rho,0)&=m_um_dm_s C_N(\xi\rho)^3\frac{1}{\rho^5}
\left(\frac{4\pi^2}{g^2}\right)^{2N}
e^{-8\pi^2/g^2(\Lambda)}\nonumber\\
\label{Instantonzero}
\end{eqnarray}
with $\xi= 1.33876$ and $C_N=0.097163$ for $N=3$.
We use $\Lambda=2.5$ GeV and $N=3$ where  $\alpha_{\rm c}(\Lambda=2.5~{\rm GeV})\simeq 0.28\pm 0.01$ \cite{Hinchliffe02}
\begin{eqnarray}
\alpha_{\rm c}(\mu)&=\frac{g_{\rm c}^2(\mu)}{4\pi}\simeq  \frac{4\pi}{\beta_0\ln(\mu^2/\Lambda_{\rm QCD}^2)}\left[1
-\frac{2\beta_1}{\beta_0^2}\frac{\ln[\ln(\mu^2/\Lambda_{\rm QCD}^2)]}{\ln(\mu^2/\Lambda_{\rm QCD}^2)}\right.\nonumber\\
&+\frac{4\beta_1^2}{\beta_0^4 \ln^2(\mu^2/\Lambda_{\rm QCD}^2)} \times
\Big(\left(\ln\Big[\ln(\mu^2/\Lambda_{\rm QCD}^2)\Big]
-\frac12\right)^2\nonumber\\
&
\left.\left.
+\frac{\beta_2\beta_0}{8\beta_1^2}-\frac54\right)\right]
\end{eqnarray}
where $\beta_0=11-\frac23 N_f,\beta_1=51-\frac{19}3 N_f$, and $\beta_2=2857-\frac{5033}9N_f+\frac{325}{27}N_f^2$. The leading factors are $m_um_dm_s T$ and $(\Lambda/T)^{ \frac{1}{3}(11N- 2N_f)}$. So, we parametrize the instanton size integration of (\ref{InstantonCal}) at $T=T_{\rm GeV}$ GeV near 1 GeV (from 700 MeV to 1.3 GeV) as
\begin{equation}
V(\theta)=-C(T)\cos(\theta),
\end{equation}
where $\theta=a/F_a$ and we parametrize $C(T)$ as
\begin{equation}
C(T)=\alpha_{\rm inst}{\rm GeV}^4 (T_{\rm GeV})^{-n}.
\end{equation}

For $\Lambda_{\rm QCD}=380 ~(440,~ 320) $ MeV \cite{Hinchliffe02}, we obtain $\alpha_{\rm inst} =  3.964\times 10^{-12} (1.274\times 10^{-11}, 9.967\times10^{-13})$, and $n=6.878~(6.789,~6.967)$.
It starts rolling down the axion potential hill when the axion mass term becomes dominant over the Hubble friction term in the equation of motion, i.e. $3H\approx m(T)$.
The Hubble parameter is $1.66 g_*^{1/2}(T) T^2/M_{\rm P}$. At $T\approx 1\textrm{GeV}$, $u,d,s,e,\mu,3\nu\bar\nu,8G,\gamma$ remain relativistic. So $g_*(T\approx 1\textrm{GeV})=(8+1)\times2+\frac{7}{8}\{
(3\times3+1+1)\times2+3\}\times2=61.75$.
Equating $3H(T)$ and $m(T)=\sqrt{C(T)/F_a^2}$, we obtain
\begin{equation}
T_\textrm{1,GeV}=\left\{
\begin{array}{l}
0.808 \times \biggl(\frac{F_{a,{\rm GeV}}}{10^{12}}\biggr)^{-0.182} ,\\
0.916 \times \biggl(\frac{F_{a,{\rm GeV}}}{10^{12}}\biggr)^{-0.184} ,\\
1.020 \times \biggl(\frac{F_{a,{\rm GeV}}}{10^{12}}\biggr)^{-0.185} ,\end{array}\right.\label{axion:roll}
\end{equation}
for $\Lambda_{\rm QCD}=320$ MeV, 380 MeV, and 440 MeV, respectively.

\section{Anharmonic effect on axion number}
In early studies \cite{Preskill83}, the anharmonic higher order terms in the axion potential were neglected and only the harmonic term (the mass term) was considered. In such an approximation, the axion total number in the comoving volume is conserved whenever the adiabaticity conditions, $H, \dot{m_a}/m_a \ll m_a$ are satisfied. Turner \cite{Steinhardt83, Turner86} considered the effect due to the anharmonic terms on the axion CDM energy density. Later, Lyth presented an extensive study on this issue \cite{Lyth91}. In the early epoch after the classical axion field starts to roll down, the axion number is not conserved because the anharmonic terms are not negligible when $\theta_1 \gtrsim 1$. Physically, the introduction of the anharmonic terms mean the axion number changing interactions so that the axion number conservation is not guaranteed.

\subsection{Adiabatic invariant}
\begin{figure}[!h]
\resizebox{0.95\columnwidth}{!}
{\includegraphics{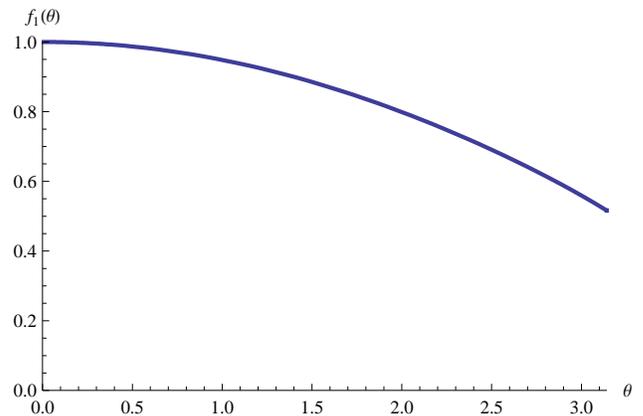}}
\caption{The correction factor from anharmonic effect.}
\label{fig:anharmonic}
\end{figure}

Here, we present the method to treat this anharmonic effect. The axion field satisfies
\begin{equation}
\ddot{\theta}+3H\dot{\theta}+{\cal V}'=0
\end{equation}
where $\theta=a/F_a, {\cal V}(\theta)= V(\theta)/F_a^2$, and the prime denotes the derivative with respect to $\theta$.
This equation of motion can be derived from the time-varying Lagrangian,
\begin{equation}
L = R^3\Big(\frac{1}{2}F_a^2\dot{\theta}^2-{ V}\Big),
\end{equation}
where $V=m_a^2F_a^2(1-\cos\theta)$. We treat here $R$ and $m_a$ as time-varying parameters. Under the adiabatic condition, $H, \dot{m_a}/m_a \ll m_a$, we have the adiabatically invariant quantity \cite{Landau},
\begin{equation}
I\equiv \frac{1}{2\pi}\oint pdq,
\end{equation}
where $q=\theta$ and $p=R^3 \dot{\theta}$ which is a conjugate momentum of $q$. Using this quantity, we can derive an invariant quantity of the axion evolution,
\begin{equation}
R^3 m_a \theta^2 f_1(\theta) = \rm constant,
\end{equation}
where
\begin{equation}
f_1(\theta)=\frac{2\sqrt{2}}{\pi\theta^2} \int^{+\theta}_{-\theta}\sqrt{\cos\theta'-\cos\theta} \; d\theta'.\label{anharmonic}
\end{equation}
The numerical result is shown in Fig. \ref{fig:anharmonic}. The same form was also obtained in \cite{Lyth91}.

Thus, we obtain
\begin{equation}
\rho_a(T_{\gamma})\simeq m_a(T_{\gamma})m_a(T_1)\biggl(\frac{R^3(T_1)
}{R^3(T_{\gamma})}\biggr)\theta_1^2 f_1(\theta_1),
\end{equation}
where $T_{\gamma}=2.73 \rm K$. We approximate here $f(\theta(T_{\gamma}))\simeq1$ since axion amplitude decreases as the universe expands and present axion amplitude is small enough to be harmonic.

\subsection{Initial anharmonic correction}
There exists the adjustment in the initial stage of the rolling down. This initial anharmonic correction will turn out to be larger than the general correction presented in the previous subsection. We assumed two adiabaticity conditions, $\dot m_a /m_a \ll m_a$ and $H\ll m_a$. However, since the expansion rate during the first few oscillations are not so small, $3 H\sim m_a$, we should be careful in treating the anharmonic correction in the initial stage. Actually, even for the case of the harmonic limit, $\theta_1 \ll 1$, the initial adjustment is needed. For a large initial misalignment angle, $\theta_1\sim1$, the duration time during which the expansion rate is not so small is longer than for the case of a small $\theta_1$. The effect of the initial adjustment is larger for the larger $\theta_1$.

\begin{figure}[!h]
\resizebox{0.95\columnwidth}{!}
{\includegraphics{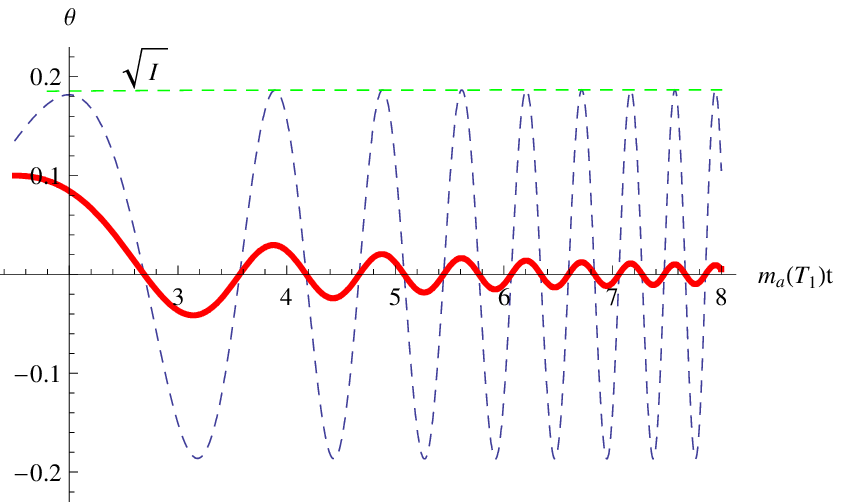}}
\vskip 0.5cm
{\includegraphics{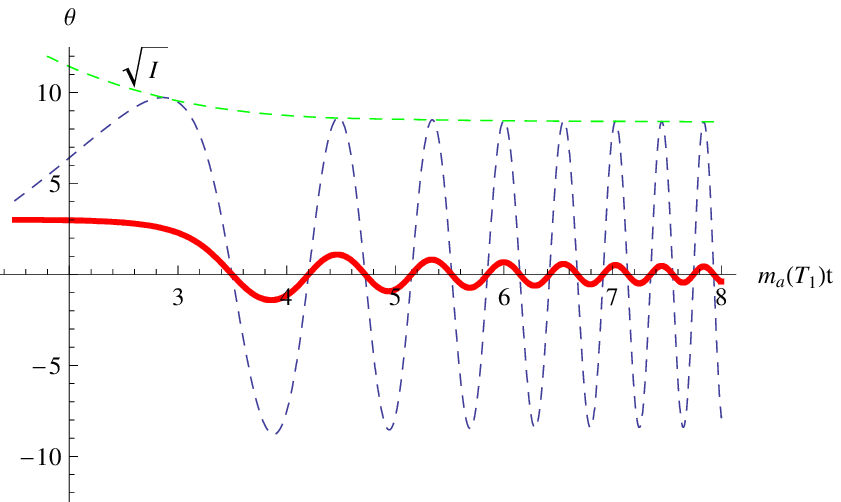}}
\caption{The solid curves represent the $\theta$ evolution for $\theta_1=0.1$ (in the upper figure) and $\theta_1=3$ (in the lower figure). The common tangents to the dashed curves are  the square root of the invariant $\sqrt{I}$. After a half period, they almost approach to the condition $H\ll m_a$.}\label{fig:thetaevolve}
\end{figure}

To estimate the initial adjustment correction, we solve the equation of motion numerically and obtain {\it  another temperature $T_2$ and the corresponding misalignment angle $\theta_2$} to use them in the subsequent calculation rather than using the original $T_1$ and $\theta_1$. We have noted that a consideration of the first half oscillation is enough for this purpose toward obtaining $H\ll m_a$ as depicted in Fig. \ref{fig:thetaevolve}. The common tangent (the dashed line) to the dashed curves is the square root of the invariant $\sqrt{I}$. In fact, this invariant shows an overshoot (a factor of 1.8 which will be shown later) from the initial misalignment angle $\theta_1$, the initial value of the solid curve, which occurs because in this initial stage the expansion rate is non-negligible compared to $m_a$. Hence we obtain a larger axion CDM energy compared to the previous estimates.

\begin{figure}[!h]
\resizebox{0.95\columnwidth}{!}
{\includegraphics{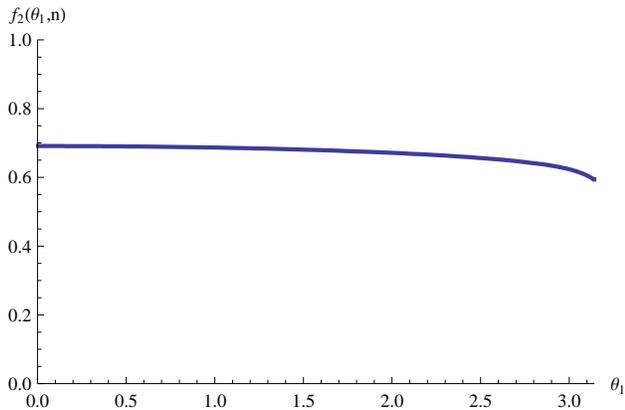}}\vskip 0.5cm
{\includegraphics{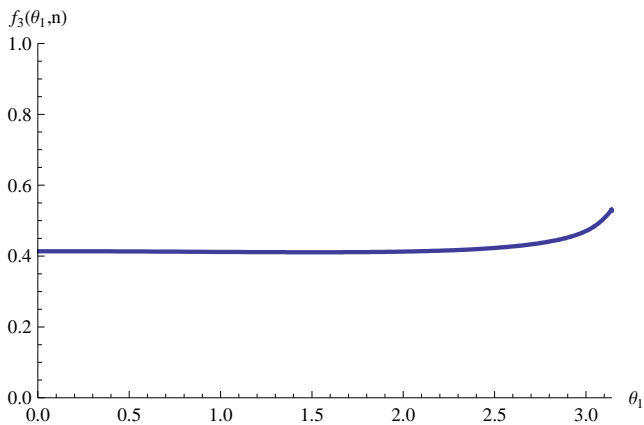}}
\caption{The initial correction factors $f_2$ and $f_3$.}\label{fig:incorrfactor}
\end{figure}

For the presentation of the result, we parametrize ${t_2}/{t_1}=f_2^{-2}(\theta_1,n)$ and ${\theta_2}/{\theta_1}=f_3(\theta_1,n)$ where $n$ is the exponent in the temperature dependent axion mass discussed earlier. The correction factors $f_2(\theta_1,n)$ and $f_3(\theta_1,n)$ are shown in Fig. \ref{fig:incorrfactor}.

If the significant entropy production is absent during the cooling period from $T_1$ down to $T_2$, $T_2/T_1=(t_2/t_1)^{-1/2}=f_2(\theta_1,n)$. To split the pure anharmonic effect from the rest, we parameterize these as $T_2/T_1=c_2 \tilde f_2(\theta_1,n)$ and $\theta_2/\theta_1=c_3\tilde f_3(\theta_1,n)$ where $c_2=\lim_{\theta_1\rightarrow0}f_2\simeq0.691$ and  $c_3=\lim_{\theta_1\rightarrow0}f_3\simeq0.414$.

\begin{figure}[!h]
\resizebox{0.95\columnwidth}{!}
{\includegraphics{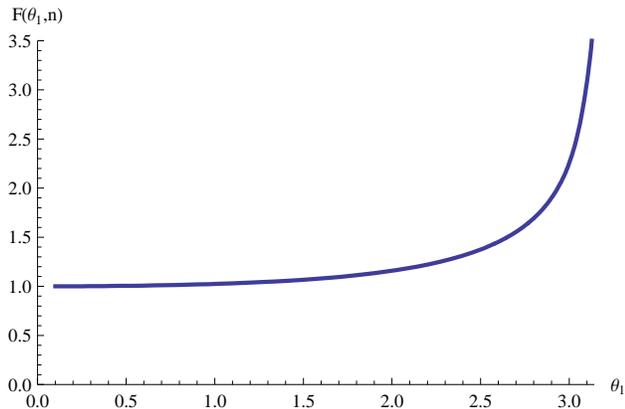}}
\caption{The combined correction factor $F(\theta_1,n)$.}\label{fig:combincorr}
\end{figure}

Then, the appropriate $\theta_1$ dependence on the relic density is given by
\begin{eqnarray}
\Omega_a &\propto& \frac{m_a(T_2)\theta_2^2}{T_2^3}f_1(\theta_2)\nonumber\\
&\propto&\frac{\theta_1^2c_3^2\tilde f_3^2(\theta_1,n)f_1(\theta_1c_3\tilde f_3(\theta_1,n))}{c_2^{3+n/2}\tilde f_2^{3+n/2}(\theta_1,n)}\nonumber\\
&=& 1.846\times\theta_1^2\times F(\theta_1,n)
\label{Fcorrfactor}
\end{eqnarray}
where the function $F(\theta_1,n)=\tilde{f}^2_3(\theta_1,n)f_1(\theta_1c_3\tilde{f}_3(\theta_1,n))/\tilde{f}_2^{3+n/2}(\theta_1,n)$ which is shown in Fig. \ref{fig:combincorr}.

\section{QCD phase transition and axion number change}

We know that the adiabatic invariant $I$ is conserved in the comoving volume. We calculated this at a later time $t_2>t_1$ where $H, \dot{m_a}/m_a \ll m_a$ are satisfied. Thus, if we know the initial misalignment angle $\theta_1=\theta(T_1)$, we can determine the current axion energy density. During the universe evolution from $T=T_1$ to $T=2.73$ K, however, there is the QCD chiral phase transition epoch when quarks and gluons in the quark-gluon phase combine each other and change to hadrons in the hadronic phase. Hence, we must check the effect of this phase transition on the total axion number. Since we do not know the exact phenomena during this phase transition, we must interpolate two regions of the quark-gluon phase and the hadronic phase as shown in Fig. \ref{fig:axmass}.

\subsection{Possible supercooling effect in phase transition}

\begin{figure}[!h]
	\resizebox{0.95\columnwidth}{!}
	{\includegraphics{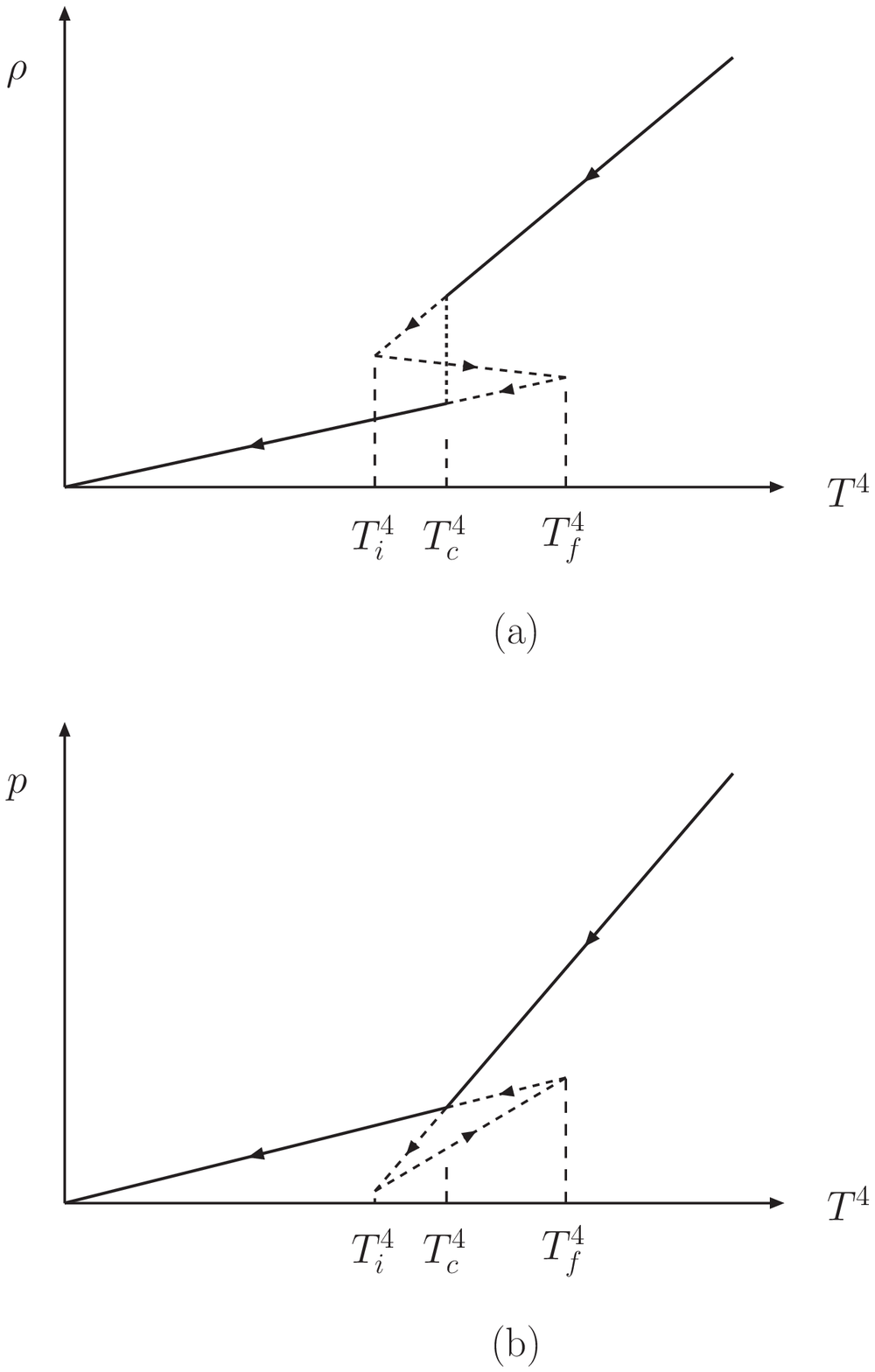}}
	\caption{The first order phase transition.}
	\label{fig:phases}
\end{figure}

\begin{figure}[!h]
\resizebox{0.95\columnwidth}{!}
{\includegraphics{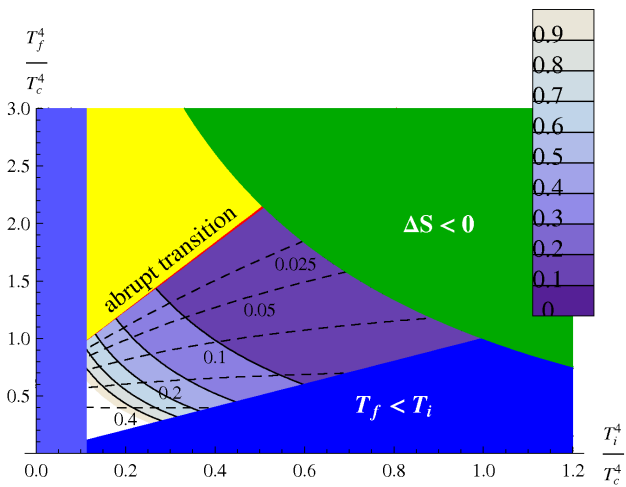}}
\caption{The physically allowed region in the $T_{\rm i}$-$T_{\rm f}$ space. Dashed lines (with numbers shown below) stand for contour of $\chi \Delta t$, and the red wedge for an exaggerated possible sudden change is $\chi \Delta t\le 10^{-3}$. The violet shaded regions show the entropy production $\Delta s/s_i$ during the phase transition.} \label{fig:temp_plot}
\end{figure}

In \cite{DeGrand:1984uq,DeGrand:1985uq}, DeGrand et al. studied supercooling effect during the QCD phase transition for various cases. We briefly explain their argument. Assuming that the universe undergoes first-order phase transition at $T\sim \Lambda_{\rm QCD}$, we shall follow bag model of phase transition. In the bag model, the equation of state is given by Fig. \ref{fig:phases}
\begin{align}
&\rho_{\rm q}(T)/B=[3r/(r-1)]\hat{T}^4 +1,\\
&p_{\rm q}(T)/B=[r/(r-1)]\hat{T}^4,\\
&T_{\rm c}s_{\rm q}(T)/B=[4r/(r-1)]\hat{T}^3,
\end{align}
in the quark-gluon phase and
\begin{align}
&\rho_{\rm h}(T)/B=3p_{\rm h}(T)/B=[3/(r-1)]\hat{T}^4,\\
&T_{\rm c}s_{\rm h}(T)/B=[4/(r-1)]\hat{T}^3,
\end{align}
in the hadronic phase, where $\hat{T}=T/T_{\rm c}$, $r=g_{\rm q}/g_{\rm h}\simeq 3$, $B=(g_{\rm h}-g_{\rm q})(\pi^2/90)T_{\rm c}^4$ and $g_{\rm q}$($g_{\rm h}$) is the relativistic degrees of freedom of quark-gluon(hadron) phase.
We shall interpolate $\rho$ and $p$ linearly between $T_{\rm i}$ and $T_{\rm f}$ like Fig. \ref{fig:phases}.  Subscript i(f) denotes the initial(final) temperature during the phase transition. After some simple algebra, we obtain the following relations \cite{DeGrand:1984uq}
\begin{align}
&R_{\rm f}/R_{\rm i}=[(\rho_{\rm i}-B)/\rho_{\rm f}]^{(\rho_{\rm f}-\rho_{\rm i})/4(\rho_{\rm f}-\rho_{\rm i}+B)},\nonumber\\
&\chi (t_{\rm f}-t_{\rm i})=\frac{1}{2} \sqrt{B}(\rho_{\rm f}-\rho_{\rm i}) \int^{\sqrt{\rho_{\rm i}}}_{\sqrt{\rho_{\rm f}}} \frac{dx}{B\rho_{\rm f}-(\rho_{\rm f}-\rho_{\rm i}+B)x^2},\nonumber\\
&\delta S/S_{\rm i}=r^{-1/4}[(\rho_{\rm i}-B)/\rho_{\rm f}]^{3B/4(\rho_{\rm f}-B-\rho_{\rm i})}\nonumber
\end{align}
where $\rho_{\rm i}=\rho_{\rm q}(T_{\rm i})$, $\rho_{\rm f}=\rho_{\rm h}(T_{\rm f})$, and $\chi=(\frac{8}{3}\pi G_NB)^{1/2} \approx 0.22 ~{\rm GeV}^2/m_{\rm Pl} $. Since we do not have enough information on $T_{\rm i}$ and $T_{\rm f}$, we treat these as free parameters. Then, we obtain the phase diagram in the $T_{\rm i}$-$T_{\rm f}$ plane as shown in Fig. \ref{fig:temp_plot}.

Parameters in colored region are not allowed because of some physical constraints. The dark blue region is excluded from the condition that $T_{\rm f}$ is equal or larger than $T_{\rm i}$. The green region is excluded from the condition that entropy never decreases. The yellow region(beyond the red line) is excluded from the constraint that energy conservation is not violated. The light blue region is excluded from the constraint on the kinematical limit of the bubble nucleation rate\cite{DeGrand:1984uq}. The thin red wedge for $\chi \Delta t<10^{-3}$ is the exaggerated region (an abrupt change case is for $\chi \Delta t<10^{-4}$ which is not noticeable in the figure) where the adiabaticity condition is violated, which is explained below.

\subsection{Axion number change}

Now we  estimate axion number change in the allowed region. To do this, we shall examine if the adiabaticity condition is still valid or not in this region. Since $H$ is decreasing during the expansion of the universe, $m_a > H$ condition is satisfied after the axion field starts to roll down. We must check whether the adiabaticity condition $\dot{m_a}/m_a < m_a$ is satisfied during the phase transition.

\subsubsection{Smooth transtion}

For a smooth transition, $T_{\rm i}=T_{\rm f}=T_{\rm c}$, we have $\rho_{\rm i}=[3r/(r-1)+1]B$, $\rho_{\rm f}=3B/(r-1)$. Thus, we obtain
\begin{align}
&R_{\rm f}/R_{\rm i}=r^{1/3},\\
&\chi \Delta t = \frac{2}{3}(r-1)^{1/2}[\arctan (4r-1)^{1/2}-\arctan \sqrt{3}] \approx 0.22,\\
&\Delta S =0,
\end{align}
where $\Delta t= t_{\rm f} -t_{\rm i}$. Therefore, we obtain $\Delta t \sim 10^{10} {\rm eV}^{-1} \gg m_a^{-1}$. Thus, the adiabaticity condition is satisfied and the axion number in the comoving volume is not changed.

\subsubsection{Abrupt transition}

Along the red wedge in Fig. \ref{fig:temp_plot}, $\rho_{\rm f}=\rho_{\rm i}$. Then,
\begin{align}
&R_{\rm f}=R_{\rm i},\\
&\Delta t=0,\\
&\Delta S/S_{\rm i} = T_{\rm f}^3/rT_{\rm i}^3 -1.
\end{align}
In this case,  $\Delta t =0$ and hence we cannot sustain the adiabaticity condition. Thus, the axion number must be changed. Depending on the phase of the axion field, the axion number may be either lowered or raised by factor 0.1 or 10 \cite{DeGrand:1985uq}.

\subsubsection{Bulk region}

In the bulk region in Fig. \ref{fig:temp_plot}, we show a reliable region satisfying the adiabaticity condition. In Fig. \ref{fig:temp_plot}, we can see that very tiny part(red wedge) of the bulk region breaks the adiabaticity condition. Thus, the axion number is not changed in the allowed region except the red wedge. In Fig. \ref{fig:temp_plot}, the entropy increase is shown as violet color shades. For a small($<50\%$) entropy production during the transition in this region, the axion energy density changes because the entropy increase ratio $\gamma$ is changed. Entropy change in the bulk region is given by
\begin{equation}
\frac{\Delta s}{s_{\rm i}} = r^{-1/4}\biggl(\frac{rT^4_{\rm i}}{T^4_{\rm f}}\biggr)^{\frac{r-1}{4(rT^4_{\rm i}-T^4_{\rm f})}}-1.
\end{equation}

Thus, we conclude that the chiral phase transition does not  have any serious effect on the initial total axion number in the comoving volume.

\section{Result}

The axion energy density, using the new numbers on the current quark masses \cite{Manohar06} and including anharmonic effect and QCD phase transition effect, is estimated as
\begin{widetext}
\begin{equation}
\begin{split}
\rho_a&(T_{\gamma}=2.73 {\rm K})=m_a(T_{\gamma})n_a(T_{\gamma})f_1(\theta_2)
=\frac{\sqrt{Z}}{1+Z} m_{\pi}f_{\pi}\frac{g_{*s}(T_{\gamma})
T_{\gamma}^3}{\gamma}\frac{3\times1.66
}{\sqrt{g_{*}(T_2)}}\frac{\theta_2^2f_1(\theta_2)}{2} \frac{F_a}{M_{\rm P}}\left(\frac{T_2}{T_1}\right)^{-3-n/2}\frac{1}{T_1}
\\
&\quad\quad \simeq 0.785 \times 10^{-11} \biggl(\frac{F_{a, \rm GeV}}{10^{12}} \biggr) \frac{1}{T_{1, \rm GeV}}\frac{\theta_1^2}{\gamma} \frac{f_3^2(\theta_1,n)
f_1(\theta_1f_3(\theta_1,n))}{f_2^{3+n/2}(\theta_1,n)} (\rm eV)^4\\
&\quad\quad \simeq 1.449 \times 10^{-11} \biggl(\frac{F_{a, \rm GeV}}{10^{12}} \biggr) \frac{1}{T_{1, \rm GeV}}\frac{\theta_1^2}{\gamma} F(\theta_1,n) (\rm eV)^4,
\end{split}\label{axion:density}
\end{equation}
\end{widetext}
where we used $Z\equiv m_u/m_d = 0.5$, $m_{\pi^0}=$135.5 MeV, $f_{\pi}=$93 MeV and $g_{*s}$(present)$=$3.91, $f(\theta_1)$ is the anharmonic correction given in Eq. (\ref{anharmonic}) and $\gamma$ is the entropy increase ratio. Inserting Eqs.(\ref{axion:roll}) to (\ref{axion:density}), we obtain for $\theta_1 \ll 1$
\begin{figure}[!h]
\resizebox{0.95\columnwidth}{!}
{\includegraphics{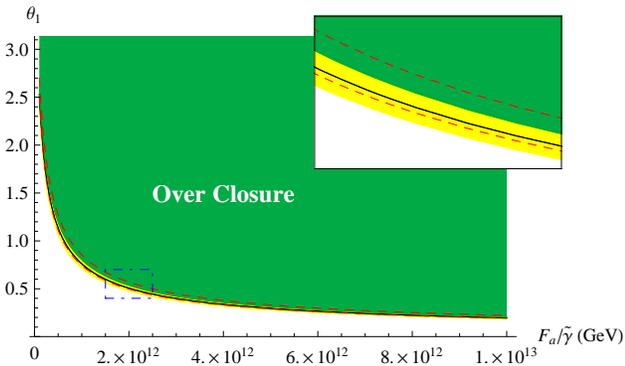}}
\caption{The bound from overclosure of the universe.  The yellow band shows the error bars of $\Lambda$ and two red dashed lines are the limits of the allowed current quark masses. The anharmonic effect is taken into account, including the initial correction factor of Eq. (\ref{Fcorrfactor}). Here, the entropy production ratio $\gamma$ is absorbed into the bracket of $F_a$: $\tilde\gamma=\gamma^{(n+4)/(n+6)}\simeq \gamma^{0.84}$.}
\label{fig:omega}
\end{figure}
\begin{equation}
\frac{\rho_a(T_{\gamma})}{({\rm eV})^4}=\left\{\begin{array}{l}
 1.794\times 10^{-11} \frac{\theta_1^2}{\gamma}F(\theta_1)
 \biggl(\frac{F_{a,\rm GeV}}{10^{12}}\biggr)^{1.182},\nonumber\\
 1.580\times 10^{-11} \frac{\theta_1^2}{\gamma}F(\theta_1)
 \biggl(\frac{F_{a,\rm GeV}}{10^{12}}\biggr)^{1.184},\nonumber\\
 1.420\times 10^{-11} \frac{\theta_1^2}{\gamma}F(\theta_1)
 \biggl(\frac{F_{a,\rm GeV}}{10^{12}}\biggr)^{1.185},
\end{array}\right.
\end{equation}
for $\Lambda_{\rm QCD}=320$ MeV, 380 MeV, and 440 MeV, respectively. For the critical density $\rho_{\rm c} = 3.978 \times 10^{-11}(h/0.701)^2$(eV)$^4$, the axion energy fraction is given by
\begin{equation}
\Omega_a \simeq \left\{\begin{array}{l}
 0.450 \biggl(\frac{\theta_1^2F(\theta_1)}{\gamma}\biggr)
\biggl(\frac{0.701}{h}\biggr)^2 \biggl(\frac{F_{a, \rm GeV}}{10^{12}} \biggr)^{1.182},\\
0.397 \biggl(\frac{\theta_1^2F(\theta_1)}{\gamma}\biggr)
\biggl(\frac{0.701}{h}\biggr)^2 \biggl(\frac{F_{a, \rm GeV}}{10^{12}} \biggr)^{1.184},\\
0.356 \biggl(\frac{\theta_1^2F(\theta_1)}{\gamma}\biggr)
\biggl(\frac{0.701}{h}\biggr)^2 \biggl(\frac{F_{a, \rm GeV}}{10^{12}} \biggr)^{1.185},
\end{array}\right.
\end{equation}
for $\Lambda_{\rm QCD}=320$ MeV, 380 MeV, and 440 MeV, respectively. From this result, we can determine the cosmological bound on the axion decay constant $F_a$ which is presented in Fig. \ref{fig:omega}. In Fig. \ref{fig:omega}, the yellow band shows the error bars of $\Lambda$ and two red lines are the allowed limits of the current quark masses, $Z$.  From the WMAP five year data \cite{Komatsu:2008hk}, $\Omega_{\rm DM} h^2 \simeq 0.1143 \pm 0.0034$, we obtain the bound $F_a \leq 10^{12} \rm GeV$ for $\theta_1 \sim O(1)$. For  $\theta_1/\sqrt{\gamma} \le 0.18$, $F_a\ge 10^{13}$ GeV is allowed. The increase of a factor of few on the upper bound of $F_a$ mainly comes from the currently used smaller current quark masses compared to the previous ones, and the overshoot of the adiabatic invariant $\sqrt{I}$ from the initial misalignment angle $\theta_1$.

\section{conclusion}

In conclusion, we presented a new expression for the cosmic axion energy density  using the new values for the current quark masses, implementing the QCD chiral symmetry breaking and the adiabatic invariant $I$. We included the effect of anharmonic terms also. Using the adiabatic invariant quantity $I$, we presented the correction function $F$ to the anharmonic effect and a factor 1.85 which arose from the overshoot from $\theta_1$.   Using the bag model for the phase transition in the early universe, the axion number is shown to be conserved in most circumstances. If the axion number change is considered during the phase transition, it is extremely fine-tuned in the $T_{\rm i}-T_{\rm f}$ space, in which case the only change is from a small entropy production during the phase transition. In this way, we obtain a factor of few larger bound on $F_a$ compared to the previous ones even for the smaller requirement on $\Omega_a$. Of course, for a small $\theta_1$ a strip with large $F_a$s is allowed, which has been used in the anthropic arguments \cite{anthropic}.

\vskip 0.5cm
\noindent {\bf Acknowledgments\ :} This work is supported in part by the Korea Research Foundation, Grant No. KRF-2005-084-C00001. KJB is supported also by the BK21 Program of the Ministry of Education and Science.

\end{document}